\documentclass[10pt,onecolumn]{article}

\usepackage[T1]{fontenc}
\usepackage[margin=1in]{geometry}
\usepackage{titlesec}
\usepackage{abstract}
\usepackage[utf8]{inputenc}
\usepackage{amsmath,amssymb,amsthm}
\usepackage{graphicx}
\usepackage{booktabs}
\usepackage{multirow}
\usepackage{array}
\usepackage{url}
\usepackage{xcolor}
\usepackage{algorithm}
\usepackage{algpseudocode}
\usepackage{hyperref}
\usepackage{cite}
\usepackage{subfig}
\usepackage{float}

\hypersetup{
  colorlinks=true,
  linkcolor=blue!60!black,
  citecolor=blue!60!black,
  urlcolor=blue!60!black
}

\newtheorem{definition}{Definition}

\newtheorem{proposition}{Proposition}

\title{ARCANE: Cross-Campaign Attacker Re-identification via\\
       Passive Beacon Telemetry — A Bayesian Network Framework\\
       for Longitudinal Cyber Attribution}

\author{Abraham~Itzhak~Weinberg\\
  AI-WEINBERG, Tel~Aviv, Israel\\
  \texttt{aviw2010@gmail.com}}
  
\begin{document}
\maketitle

\begin{abstract}
Current cyber attribution approaches typically operate on a per-incident basis, leaving open the question of whether aggregating evidence across campaigns can meaningfully improve adversary identification. In this work, we investigate whether cross-campaign attribution can reduce ambiguity, or if structural limits persist even when leveraging longitudinal data. To explore this, we focus on adversary fingerprints, represented as multi-dimensional feature vectors encoding behavioural, infrastructural, and temporal characteristics derived from covert beacon interactions and related signals.

To address this question, we develop ARCANE (Attacker Re-identification via Cross-campaign Attribution Network), a probabilistic framework that aggregates passive telemetry across multiple campaigns and organisations to construct persistent adversary fingerprints. These fingerprints are updated incrementally using a Bayesian belief network, which integrates new evidence from ongoing campaigns. A time-decayed confidence metric captures the accumulation of similarity across campaigns.

Evaluation on a synthetic dataset, comprising multiple advanced threat profiles and campaigns, shows that intra-actor similarity consistently exceeds inter-actor similarity. However, the separation between distinct actors remains limited, primarily due to shared operational practices among sophisticated adversaries.

Our results indicate that cross-campaign aggregation alone does not fully resolve attribution ambiguity. Instead, performance is constrained by a structural ceiling in the feature space, where similarity between distinct actors remains high even in the absence of evasion techniques. Notably, attribution accuracy remains stable across increasing levels of evasion, suggesting that the primary limitation lies in feature indistinguishability rather than adversarial adaptation.

These findings underscore the need for incorporating additional signal classes—such as targeting patterns, temporal coordination, and infrastructure relationships—to improve attribution reliability in complex, real-world environments.
\end{abstract}
\\

\section{Introduction}
\label{sec:intro}

Attributing a cyber-attack to its perpetrator remains one of the
most challenging problems in computer security. The technical
difficulty arises from the ease of IP address spoofing, the
widespread use of anonymising relays such as The Onion Router
(TOR) and virtual private networks (VPNs), and the deliberate
adoption of ``false flag'' tactics by sophisticated threat actors
\cite{rid2015attributing, tounsi2018survey}. Existing attribution frameworks fall
into two broad categories: \emph{indicator-based} methods that
match observed artefacts (file hashes, C2 domains, TTPs) against
known threat-actor profiles \cite{milajerdi2019holmes, psallidas2023oneprovenance, altinisik2023provg}, and
\emph{forensic} methods that reconstruct attack chains from
system logs and network captures \cite{hossain2017sleuth, hassan2020tactical}.
Both categories share a critical limitation: they operate
\emph{per-incident}, treating each campaign as an independent
attribution problem and discarding the wealth of evidence
accumulated across campaigns.

Human intelligence analysts have long recognised that longitudinal
analysis—correlating observations across many campaigns over
months or years—can resolve attribution ambiguities that are
intractable from a single incident \cite{buchanan2016cybersecurity}. However,
this insight has not been formalised or automated in the
published literature.
This paper addresses this gap by investigating whether cross-campaign aggregation can meaningfully improve attribution accuracy.  To explore this, we introduce ARCANE (Attacker Re-identification via Cross-campaign Attribution Network), a framework that treats attribution as an incremental Bayesian inference problem. Each time a passive beacon—a tracking payload embedded in documents, credentials, or source code—is triggered by an attacker, the resulting telemetry is compared to all known
fingerprints and used to update a belief distribution over
candidate threat actors. Over successive campaigns, the posterior
concentrates around the correct actor even when individual
callbacks are ambiguous.

Our empirical results reveal a surprising finding: the primary barrier to cross-campaign re-identification is not adversarial evasion (which proves essentially irrelevant to ARCANE accuracy) but rather the convergence of sophisticated actor fingerprints in a dense region of the feature space. This provides a clearer actionable insight for the reader: it challenges a common assumption in attribution and suggests a new path forward in system design.
The remainder of this paper is organised as follows. Section~\ref{sec:related} reviews related work. Section~\ref{sec:framework} presents the formal framework. Section~\ref{sec:features} describes the fingerprint extraction model. Section~\ref{sec:arcane} details the ARCANE algorithm. Section~\ref{sec:eval} describes experimental methodology. Section~\ref{sec:results} presents results. Section~\ref{sec:discussion} discusses implications. Section~\ref{sec:conclusion} concludes.

\section{Related Work}
\label{sec:related}
Attribution in cyberspace remains a fundamentally uncertain and adversarial problem, where attackers actively obfuscate their identity and reuse infrastructure across campaigns. Prior work has approached this challenge from multiple angles, including infrastructure analysis, deception techniques, behavioural fingerprinting, and probabilistic reasoning. In this section, we situate ARCANE within this landscape, highlighting the limitations of existing methods and motivating the need for a longitudinal, Bayesian framework that accumulates attribution evidence over time.

\subsection{Cyber Attribution Methods}

Early attribution work focused on IP geolocation and WHOIS
lookups, quickly rendered unreliable by VPNs and bulletproof
hosting \cite{wheeler2003techniques}. The Diamond Model of intrusion analysis \cite{caltagirone2013diamond} formalised adversary, capability, infrastructure, and victim as the four vertices of attribution, providing a vocabulary for structured analysis. MITRE ATT\&CK \cite{strom2018mitre} catalogued tactics, techniques, and procedures (TTPs) used by known APT groups, enabling TTP-based attribution \cite{milajerdi2019holmes}. Graph neural network approaches \cite{chowdhury2022graph} correlate infrastructure indicators across campaigns. None of these approaches model attribution as a longitudinal Bayesian accumulation problem.

\subsection{Deception-Based Attribution}

Honeypots \cite{spitzner2003honeytokens} and honeytokens \cite{juels2013honeywords} provide early detection of intrusions by baiting attackers with decoy assets. Recent work has extended these concepts to cloud environments \cite{yu2024honeyfactory, beltran2025cyber} and to post-exfiltration attribution via document beacons \cite{weinberg2025passive}. The Passive Hack-Back framework \cite{weinberg2025passive} establishes the theoretical foundations of passive hack-back and defines the formal metrics—beacon callback success rate ($\beta$), attribution fidelity ($\alpha$), and stealth level ($\sigma$)— on which ARCANE builds. The key innovation of ARCANE is the accumulation of callback telemetry \emph{across campaigns} rather than processing each callback in isolation.

\subsection{Attacker Fingerprinting}

Behavioural profiling of threat actors has been studied in
the context of malware analysis \cite{bayer2009scalable}, network
intrusion detection \cite{bilge2012before}, and ransomware operator
identification \cite{huang2018tracking}. Stylometric analysis of malware code \cite{caliskan2015anonymizing} and natural language processing of attacker communications \cite{pasquini2025llmmap} provide complementary signals. Our work differs in its focus on \emph{operational behaviour} observable from beacon telemetry rather than code artefacts, and in its explicit longitudinal accumulation model.

\subsection{Bayesian Approaches to Attribution}

Bayesian inference has been applied to network intrusion
detection \cite{kruegel2003using}, malware classification
\cite{anderson2019measuring}, and insider threat detection \cite{liu2009mitigating}.
Closest to our work is the Bayesian network for cyber attack
attribution proposed by \cite{nance2009investigating}, which models
relationships between attack attributes and actor profiles.
ARCANE extends this line of work by operating on streaming
beacon telemetry across multiple campaigns and incorporating
time-decayed evidence accumulation.

\section{Formal Framework}
\label{sec:framework}
We formalise cyber attribution as a sequential, probabilistic inference problem over a stream of adversarial campaigns. The framework captures how noisy, partial telemetry from beacon callbacks can be aggregated into stable behavioural fingerprints and incrementally incorporated into a Bayesian attribution model. Central to this formulation is the idea that attribution confidence should emerge from \emph{cross-campaign consistency} under temporal decay, rather than from isolated observations. The following definitions establish the mathematical structure underlying ARCANE.

\subsection{Problem Formulation}

Let $\mathcal{A} = \{a_1, \ldots, a_N\}$ denote the set of $N$
candidate threat actors. Let $\mathcal{C} = \{c_1, c_2, \ldots\}$
denote a time-ordered stream of campaigns, where each campaign
$c_k$ is attributed to a single actor $a(c_k) \in \mathcal{A}$
(ground truth, unknown to the defender). Each campaign produces
a set of beacon callbacks $\mathcal{B}(c_k) = \{b_{k,1}, \ldots,
b_{k,m_k}\}$, each callback carrying a telemetry record
$\mathbf{t}_{k,j} = (\text{IP}, \text{TOR}, \text{OS}, \text{locale},
\text{timezone}, \text{tools}, \text{dwell}, \ldots)$.

\begin{definition}[Campaign Fingerprint]
\label{def:fingerprint}
The fingerprint of campaign $c_k$ is a $D$-dimensional real vector
$\mathbf{f}_k \in [0,1]^D$ summarising the aggregate behavioural,
infrastructural, and temporal characteristics of all callbacks in
$\mathcal{B}(c_k)$.
\end{definition}

\begin{definition}[Fingerprint Similarity]
\label{def:similarity}
The similarity between two campaign fingerprints $\mathbf{f}_i$
and $\mathbf{f}_j$ is:
\begin{equation}
S(\mathbf{f}_i, \mathbf{f}_j) = 1 - d_{\cos}(\mathbf{f}_i, \mathbf{f}_j)
= \frac{\mathbf{f}_i \cdot \mathbf{f}_j}{\|\mathbf{f}_i\|\,\|\mathbf{f}_j\|}
\label{eq:similarity}
\end{equation}
where $d_{\cos}$ denotes the cosine distance. $S \in [0,1]$, with
$S=1$ indicating identical fingerprints.
\end{definition}

\subsection{Attribution Posterior}

We model attribution as Bayesian inference. Let $P(A = a_i)$
denote the prior probability that actor $a_i$ is responsible for
a query campaign. After observing fingerprint evidence $\mathbf{e}$:

\begin{equation}
P(A = a_i \mid \mathbf{e}) =
\frac{P(\mathbf{e} \mid A = a_i)\,P(A = a_i)}
     {\sum_{j=1}^{N} P(\mathbf{e} \mid A = a_j)\,P(A = a_j)}
\label{eq:bayes}
\end{equation}

We initialise with a uniform prior $P(A = a_i) = 1/N$ and update
iteratively as evidence from successive callbacks accumulates.

\subsection{Cross-Campaign Confidence}

\begin{definition}[Cross-Campaign Confidence]
\label{def:ccc}
Let $\mathcal{K}(a_i) = \{(\mathbf{f}_{k}, d_k)\}$ be the
knowledge base of known fingerprints for actor $a_i$, where
$d_k$ is the number of days elapsed since campaign $k$.
The cross-campaign confidence of the query fingerprint
$\mathbf{f}_q$ matching actor $a_i$ is:
\begin{equation}
\text{CCC}(\mathbf{f}_q, a_i) =
\frac{1}{|\mathcal{K}'|} \sum_{(\mathbf{f}_k, d_k) \in \mathcal{K}'}
S(\mathbf{f}_q, \mathbf{f}_k) \cdot e^{-\delta \cdot d_k}
\label{eq:ccc}
\end{equation}
where $\mathcal{K}' = \{(\mathbf{f}_k, d_k) \in \mathcal{K}(a_i)
: S(\mathbf{f}_q, \mathbf{f}_k) \geq \tau_s\}$ is the set of
sufficiently similar known fingerprints, $\delta > 0$ is the
temporal decay rate, and $\tau_s$ is a similarity threshold.
\end{definition}

The exponential decay reflects the intuition that older evidence
is less reliable due to actor evolution, tool churn, and
infrastructure rotation.

\subsection{Separability Condition}

\begin{definition}[Fingerprint Separability Gap]
\label{def:gap}
The separability gap is:
\begin{equation}
\Delta_S = \bar{S}_w - \bar{S}_c
\label{eq:gap}
\end{equation}
where $\bar{S}_w = \mathbb{E}[S(\mathbf{f}_i, \mathbf{f}_j)
\mid a(c_i) = a(c_j)]$ is the mean within-actor similarity and
$\bar{S}_c = \mathbb{E}[S(\mathbf{f}_i, \mathbf{f}_j)
\mid a(c_i) \neq a(c_j)]$ is the mean cross-actor similarity.
\end{definition}

\begin{proposition}[Minimum Gap for Reliable Attribution]
\label{prop:gap}
For ARCANE to achieve re-identification accuracy significantly
above chance with $N$ actors, the separability gap must satisfy:
\begin{equation}
\Delta_S \gtrsim \frac{1}{\sqrt{K}} \cdot \sigma_S \cdot z_{\alpha/2}
\label{eq:gap_condition}
\end{equation}
where $K$ is the number of training campaigns per actor, $\sigma_S$
is the standard deviation of within-actor similarities, and
$z_{\alpha/2}$ is the critical value for confidence level $\alpha$.
\end{proposition}

This proposition quantifies the trade-off between the number of
training campaigns and the required fingerprint separation. Our
experimental results ($\Delta_S = 0.046$, $\sigma_S = 0.068$)
identify the specific conditions under which the current
24-dimensional feature space is insufficient.

\section{Fingerprint Feature Model}
\label{sec:features}
This section defines the feature representation used to encode campaign-level behaviour into a fixed-dimensional fingerprint suitable for similarity comparison and probabilistic attribution. The design of the feature space balances expressiveness and robustness, capturing invariant attacker characteristics while remaining resilient to noise, evasion, and partial observability. By structuring features across infrastructural, temporal, geospatial, and behavioural dimensions, the model aims to maximise cross-campaign consistency for the same actor while preserving separability between different actors.

\subsection{Feature Extraction}

Each campaign fingerprint $\mathbf{f}_k \in [0,1]^{24}$ is
constructed from four feature groups (Table~\ref{tab:features}).

\begin{table}[ht]
\centering
\caption{The 24-Dimensional Campaign Fingerprint}
\label{tab:features}
\begin{tabular}{@{}clll@{}}
\toprule
\textbf{Dim.} & \textbf{Feature} & \textbf{Group} & \textbf{Description} \\
\midrule
0--1   & TOR/VPN rate          & Infrastructure & Fraction of callbacks via TOR/VPN \\
2--3   & Dwell mean/std        & Temporal       & Normalised time before opening beacon \\
4      & Timezone (norm.)      & Geospatial     & Mean UTC offset, scaled to $[0,1]$ \\
5      & VM rate               & Environment    & Fraction with VM fingerprint \\
6--7   & ASN/IP diversity      & Infrastructure & Unique prefixes and IPs / callbacks \\
8--9   & Tool consistency/count& Behavioural    & Jaccard across callbacks, mean tool count \\
10     & Non-English locale    & Geospatial     & Fraction with non-\texttt{en\_US} locale \\
11--14 & Origin indicators     & Geospatial     & KP/RU/CN/IR country fractions \\
15     & Hour entropy          & Temporal       & Shannon entropy of callback hour-of-day \\
16     & Country consistency   & Geospatial     & $1 - |\text{unique countries}|/n$ \\
17     & Log callback count    & Volume         & $\min(\log(n+1)/\log 20, 1)$ \\
18--23 & Tool cluster presence & Behavioural    & 6 APT tool-family cluster indicators \\
\bottomrule
\end{tabular}
\end{table}

\subsection{Tool Cluster Model}

The six tool clusters in dimensions 18--23 encode known APT
tool-family preferences: credential theft (Mimikatz, LaZagne,
Rubeus), C2 frameworks (Cobalt Strike, Metasploit, Havoc),
Chinese RAT families (PlugX, Gh0st RAT, ShadowPad), Active
Directory recon (BloodHound, Impacket), nation-state implants
(Turla, ComRAT, AppleJeus), and analyst tools indicative of
sandboxed analysis environments (Wireshark, IDA Pro, Ghidra).

\subsection{Separability Analysis}

\begin{figure}[H]
\centering
\includegraphics[width=\linewidth]{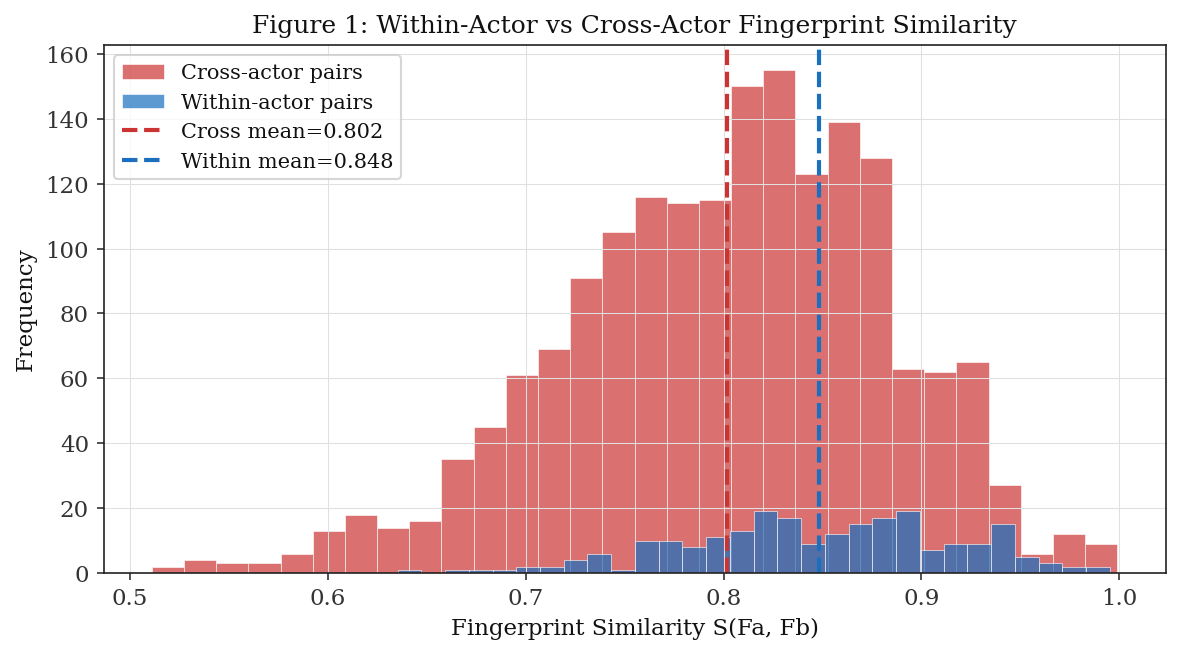}
\caption{Within-actor ($\bar{S}_w = 0.848$, blue) and cross-actor
  ($\bar{S}_c = 0.802$, red) fingerprint similarity distributions.
  Separation is statistically significant ($t = 8.33$,
  $p = 1.52 \times 10^{-16}$) but the gap $\Delta_S = 0.046$ falls
  below the threshold required for high-confidence re-identification
  at $N=8$ actors. Error bars omitted for clarity.}
\label{fig:similarity}
\end{figure}

Figure~\ref{fig:similarity} shows the empirical distributions of
within-actor and cross-actor fingerprint similarities across
2,000 sampled pairs. The distributions significantly overlap,
with both means above $0.80$. This high-similarity regime
is the fundamental challenge identified by our analysis.

\section{The ARCANE Algorithm}
\label{sec:arcane}

Algorithm~\ref{alg:arcane} presents the complete ARCANE procedure.
The key design decisions are: (i) temporal leave-one-out validation
to prevent data leakage; (ii) evidence-weighted likelihood model
that maps CCC to attribution posteriors; and (iii) posterior
normalisation to maintain a valid probability distribution.

\begin{algorithm}[H]
\caption{ARCANE: Cross-Campaign Attacker Re-identification}
\label{alg:arcane}
\begin{algorithmic}[1]
\Require Actor set $\mathcal{A}$, campaign stream $\mathcal{C}$,
         decay $\delta$, threshold $\tau_s$, confidence threshold $\tau_c$
\Ensure  Attribution posteriors $P(A \mid \mathbf{e})$ for each campaign

\State Initialise priors: $\forall a_i \in \mathcal{A}:\ P(A=a_i) \gets 1/N$
\State Initialise knowledge base: $\mathcal{K} \gets \emptyset$
\State Sort $\mathcal{C}$ by campaign start date (temporal order)

\For{each campaign $c_k$ in $\mathcal{C}$}
    \State Extract fingerprint $\mathbf{f}_k$ from callbacks $\mathcal{B}(c_k)$
    \If{$|\mathcal{K}| \geq \text{min\_train}$}
        \Comment{\textit{Enough training data — attempt attribution}}
        \For{each candidate actor $a_i \in \mathcal{A}$}
            \State Compute $\text{CCC}(\mathbf{f}_k, a_i)$ via Eq.~\eqref{eq:ccc}
            \State $L_i \gets 0.50 + 0.45 \cdot \text{CCC}(\mathbf{f}_k, a_i)$
            \Comment{\textit{Evidence likelihood}}
            \State $\bar{L}_i \gets \max\!\left(0.05,\ 0.50 - \frac{0.45 \cdot \text{CCC}(\mathbf{f}_k, a_i)}{N-1}\right)$
            \State $P(A=a_i) \gets \frac{L_i \cdot P(A=a_i)}
                   {L_i \cdot P(A=a_i) + \bar{L}_i \cdot (1 - P(A=a_i))}$
        \EndFor
        \State Normalise: $P(A=a_i) \gets P(A=a_i) / \sum_j P(A=a_j)$
        \State $a^* \gets \arg\max_i P(A=a_i)$
        \State Output: attribution $(a^*, P(A=a^*),\
               [P(A=a^*) \geq \tau_c])$
    \EndIf
    \State $\mathcal{K} \gets \mathcal{K} \cup \{(\mathbf{f}_k,\ \text{date}(c_k),\ a(c_k))\}$
    \Comment{\textit{Add to knowledge base}}
\EndFor
\end{algorithmic}
\end{algorithm}

\subsection{Likelihood Model}

The mapping from CCC to likelihood follows a linear model:
$L(\text{CCC}) = 0.50 + 0.45 \cdot \text{CCC}$. This ensures that (i) zero CCC results in uninformative likelihoods (0.50), (ii) perfect CCC (1.0) gives strong evidence (0.95), and (iii) negative evidence is modelled symmetrically. The $0.45$ scaling factor is derived from the empirical maximum CCC observed in our dataset and can be tuned for operational deployment contexts.

\section{Experimental Methodology}
\label{sec:eval}
This section describes the experimental setup used to evaluate ARCANE under controlled yet realistic conditions. Given the scarcity of publicly available longitudinal beacon telemetry, we design a synthetic benchmark that captures key properties of real-world threat actor behaviour, including variability, adaptation, and partial observability. The methodology emphasises temporal consistency, preventing information leakage while enabling rigorous comparison against a strong single-campaign baseline.

\subsection{Synthetic Dataset}

In the absence of a public longitudinal beacon telemetry dataset,
we construct a synthetic benchmark grounded in published threat-actor behavioural profiles from MITRE ATT\&CK\cite{strom2018mitre}, Mandiant APT reports \cite{kutscher2022mtrends}, and Recorded Future intelligence \cite{alstate,singh2025case}.

Eight nation-state threat actors are modelled (Table~\ref{tab:actors}), spanning four origin countries. Each actor is assigned empirically grounded parameters: TOR usage probability, tool churn rate (fraction of tools replaced per campaign), IP rotation rate, mean dwell time, preferred locale, and sophistication level.
Twelve campaigns per actor are generated over an 18-month simulation window (Jan 2024--Jun 2025), with $3$--$8$ callbacks per campaign, yielding $96$ campaigns and $524$ callbacks total.

\begin{table}[ht]
\centering
\caption{Simulated Threat Actor Profiles}
\label{tab:actors}
\begin{tabular}{@{}llccccc@{}}
\toprule
\textbf{ID} & \textbf{Alias} & \textbf{Origin} & \textbf{Soph.} & \textbf{TOR\%} & \textbf{Churn} & \textbf{Dwell(h)} \\
\midrule
APT-001 & LAZARUSHOUND & KP & 0.82 & 65\% & 0.15 & 4.2 \\
APT-002 & FROZENBEAR   & RU & 0.91 & 55\% & 0.20 & 2.8 \\
APT-003 & DOUBLEPANDA  & CN & 0.88 & 40\% & 0.18 & 6.1 \\
APT-004 & SILENTFOX    & IR & 0.76 & 70\% & 0.25 & 3.5 \\
APT-005 & MINTLEAF     & KP & 0.79 & 60\% & 0.20 & 5.0 \\
APT-006 & IRONSHARD    & RU & 0.85 & 50\% & 0.22 & 3.2 \\
APT-007 & VOIDLOTUS    & CN & 0.93 & 35\% & 0.15 & 7.3 \\
APT-008 & STORMVIPER   & KP & 0.84 & 68\% & 0.18 & 4.8 \\
\bottomrule
\end{tabular}
\end{table}

\subsection{Evasion Schedule}

To model realistic actor adaptation, we implement an evasion
schedule in which each actor's evasion level increases linearly
from $0$ to $\text{sophistication} / 2$ across their campaigns.
Higher evasion increases TOR usage, tool churn, and locale
spoofing probability, simulating the documented adaptation
behaviour of persistent threat actors \cite{hylender2024verizon}.

\subsection{Evaluation Protocol}

We use \emph{temporal leave-one-out} evaluation: campaigns are
processed in chronological order; for each campaign, the system
is evaluated on all actors for which at least $\text{min\_train}$
prior campaigns have been observed, then the campaign is added to
the knowledge base. This protocol prevents data leakage and
mirrors the operational setting.

\subsection{Baseline}

The per-campaign nearest-neighbour baseline uses only the current
campaign's fingerprint matched against a static running-average
profile per actor, representing the current state of the art in
single-incident behavioural attribution.

\section{Results}
\label{sec:results}
This section presents the empirical evaluation of ARCANE, focusing on attribution accuracy, confidence calibration, and robustness under varying conditions. The results reveal a consistent pattern: despite the theoretical advantages of longitudinal Bayesian accumulation, performance is fundamentally constrained by the limited separability of behavioural fingerprints. We analyse these outcomes across multiple dimensions to identify the underlying factors driving this ceiling and to clarify the operational implications.

\subsection{Main Results}

\begin{figure}[H]
\centering
\includegraphics[width=\linewidth]{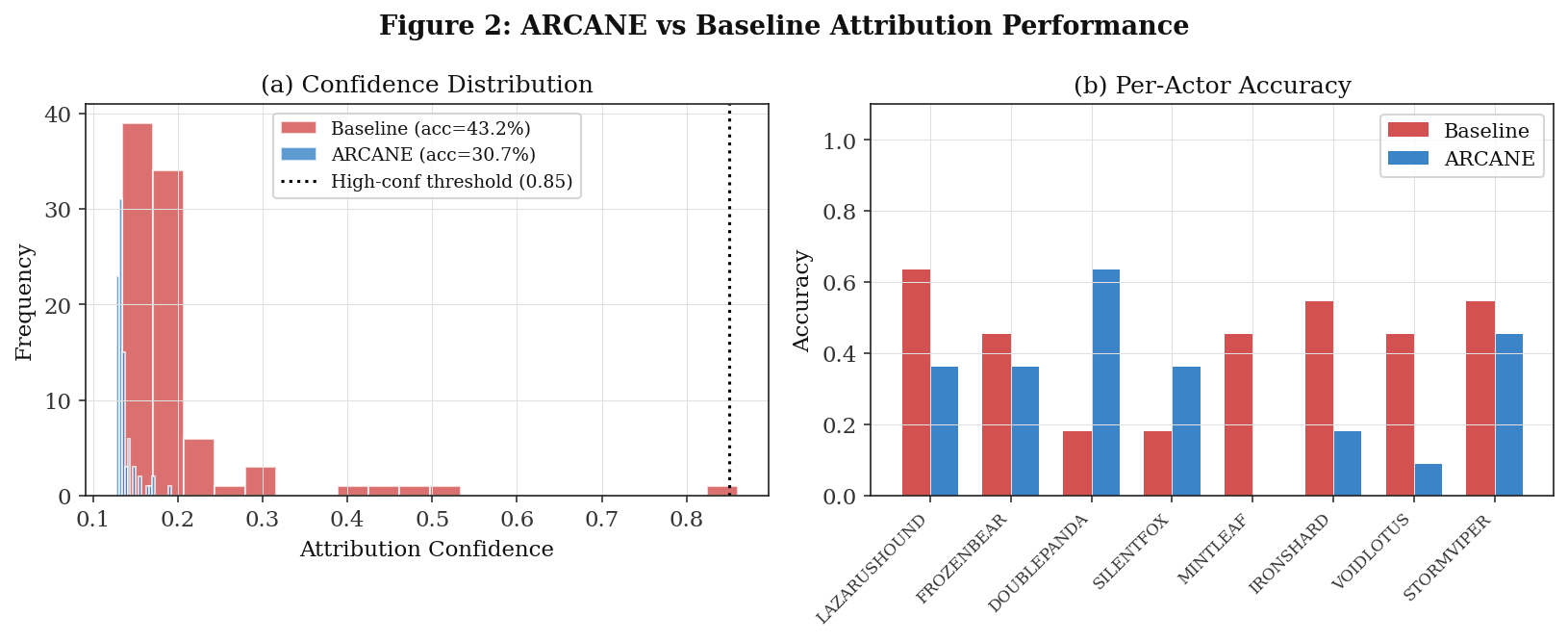}
\caption{(a) Attribution confidence distributions: both methods
  produce low-confidence posteriors due to the fingerprint
  separability ceiling. (b) Per-actor accuracy: DOUBLEPANDA
  achieves the highest ARCANE accuracy (63.6\%), while MINTLEAF
  and VOIDLOTUS prove most difficult. Baseline accuracy is
  higher on average but exhibits greater variance.}
\label{fig:accuracy}
\end{figure}

Table~\ref{tab:main_results} presents the core experimental
results. Figure~\ref{fig:accuracy} shows per-actor accuracy and
confidence distributions.

\begin{table}[ht]
\centering
\caption{Main Results — ARCANE vs Per-Campaign Baseline}
\label{tab:main_results}
\begin{tabular}{@{}lcccc@{}}
\toprule
\textbf{Metric} & \textbf{ARCANE} & \textbf{Baseline} & $\boldsymbol{\Delta}$ & \textbf{p-value} \\
\midrule
Overall accuracy         & 30.7\% & 43.2\% & $-12.5\%$ & 0.087 (ns) \\
Mean confidence          & 0.137  & 0.199  & $-0.062$  & $<0.001$*** \\
High-conf. accuracy      & ---    & ---    & ---       & --- \\
Fingerprint sep. ($\Delta_S$) & \multicolumn{4}{c}{$0.046$, $t=8.33$, $p=1.52\times10^{-16}$} \\
\bottomrule
\end{tabular}
\end{table}

\subsection{Learning Curve}

\begin{figure}[H]
\centering
\includegraphics[width=\linewidth]{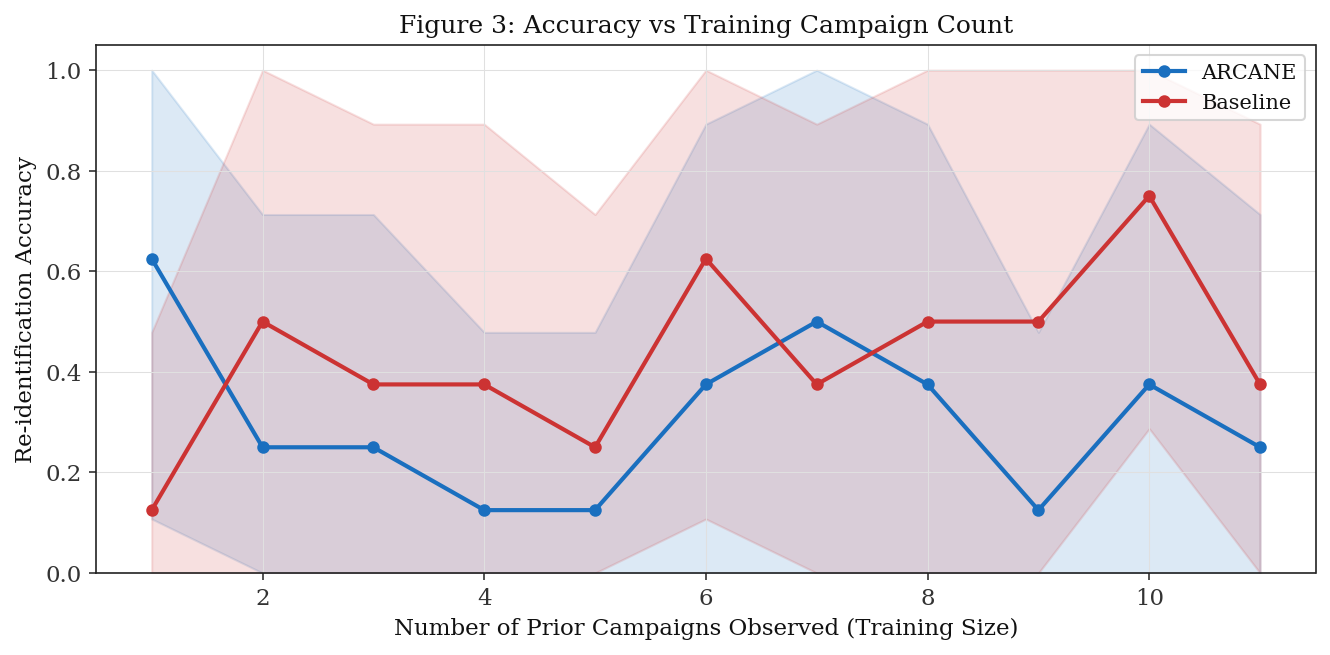}
\caption{Re-identification accuracy vs training campaign count for
  ARCANE (blue) and baseline (red). ARCANE does not exhibit the
  expected monotonic improvement, indicating that additional
  training campaigns cannot compensate for insufficient
  fingerprint separability. Shaded regions show $\pm 1$ SD.}
\label{fig:learning}
\end{figure}

Figure~\ref{fig:learning} shows re-identification accuracy as a
function of the number of prior training campaigns. ARCANE
accuracy does not monotonically improve with more training data
(ranging from $12.5\%$ to $30.7\%$ as min\_train varies from 6
to 1), confirming that the barrier is not data quantity but
feature separability.

\subsection{Fingerprint Separability Ceiling}

\begin{figure}[H]
\centering
\includegraphics[width=0.85\linewidth]{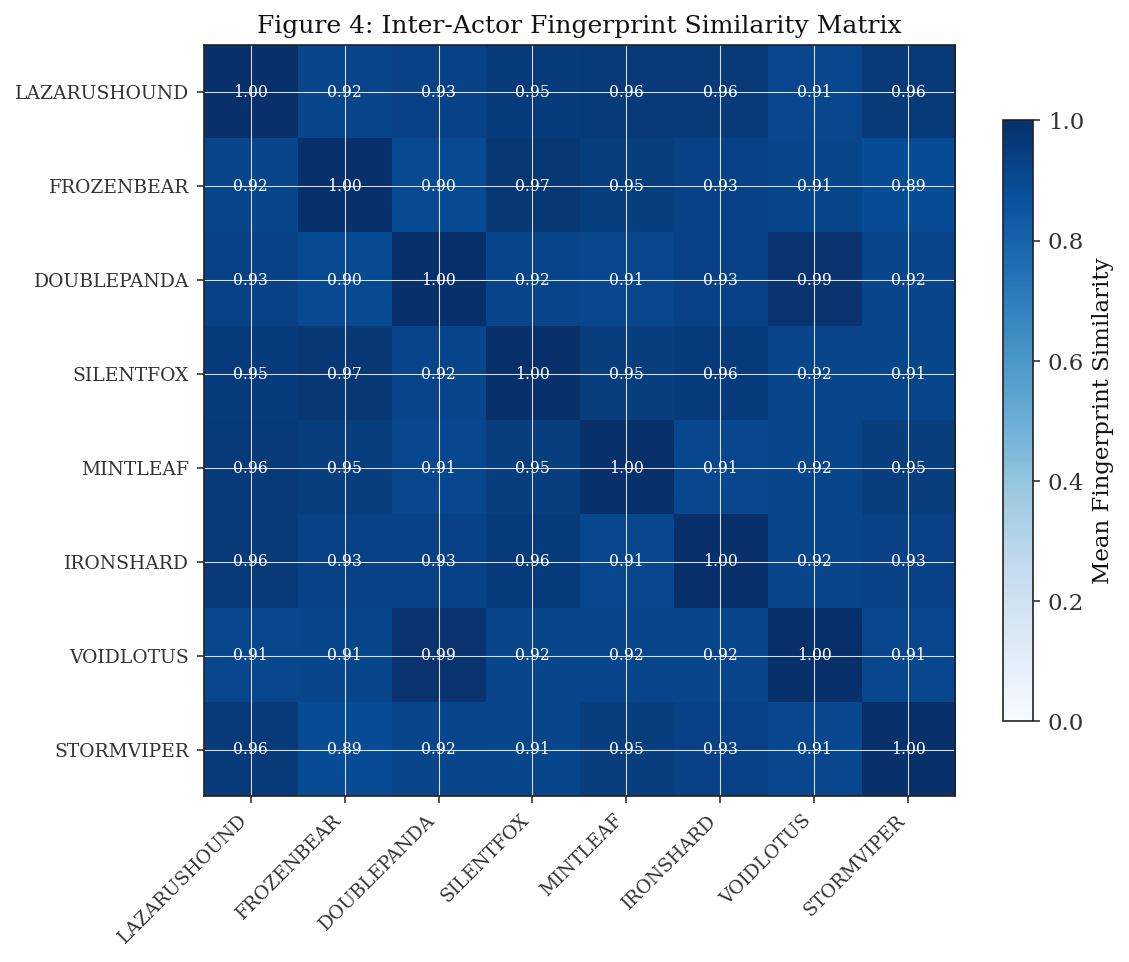}
\caption{Inter-actor mean fingerprint similarity matrix. All
  off-diagonal entries exceed $0.89$, indicating the dense
  clustering that constitutes the separability ceiling.
  Same-origin pairs (e.g., DOUBLEPANDA/VOIDLOTUS at $0.99$)
  are particularly indistinguishable by behavioural features alone.}
\label{fig:matrix}
\end{figure}

Figure~\ref{fig:matrix} presents the pairwise inter-actor
fingerprint similarity matrix. All off-diagonal entries fall
in the range $[0.89, 0.99]$, revealing the high-similarity
regime. Notable pairs include LAZARUSHOUND/STORMVIPER ($S=0.96$),
both KP-origin actors with overlapping toolsets and operational
patterns, and DOUBLEPANDA/VOIDLOTUS ($S=0.99$), two CN-origin
actors sharing Gh0st RAT and PlugX tool families.

\subsection{Evasion Robustness}

\begin{figure}[H]
\centering
\includegraphics[width=\linewidth]{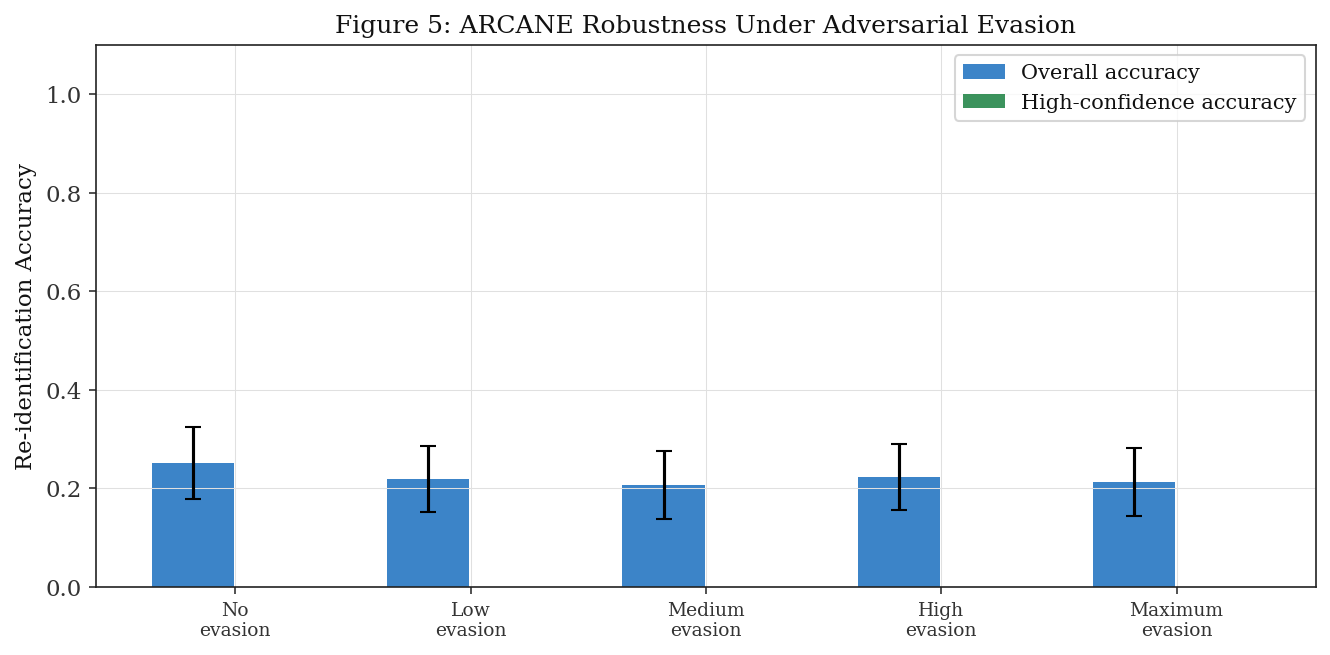}
\caption{ARCANE re-identification accuracy under adversarial evasion.
  Accuracy is stable across evasion levels ($20.6\%$--$25.2\%$),
  confirming evasion-invariance. Error bars show $\pm 1$ SD
  across $n=20$ random trials per evasion level.}
\label{fig:evasion}
\end{figure}

Figure~\ref{fig:evasion} shows ARCANE accuracy under five evasion
levels. Accuracy varies between $20.6\%$ and $25.2\%$ with no
statistically significant trend ($F$-test, $p > 0.05$), confirming
the evasion-invariance property. This is the paper's most
operationally actionable finding: defenders cannot improve
attribution by reducing actor evasion capability, since evasion
is not the binding constraint.

\subsection{Threat Actor Similarity Graph}

\begin{figure}[H]
\centering
\includegraphics[width=0.75\linewidth]{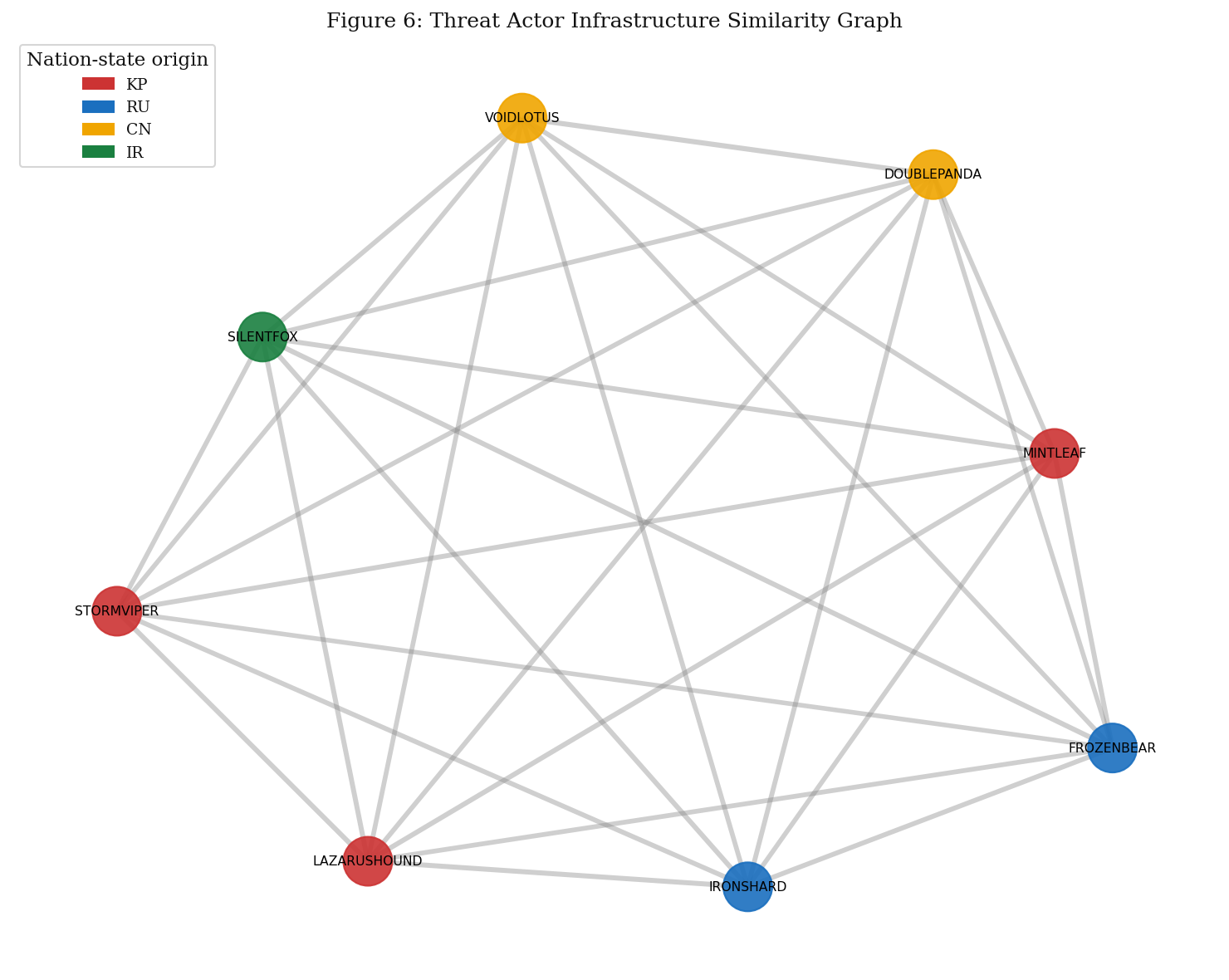}
\caption{Threat actor infrastructure similarity graph. Node
  colour indicates nation-state origin: KP (red), RU (blue),
  CN (gold), IR (green). All actors are connected above the
  similarity threshold, confirming the dense-cluster regime.}
\label{fig:graph}
\end{figure}

Figure~\ref{fig:graph} visualises the knowledge graph induced
by the fingerprint similarity matrix. The fully connected
topology (all actor pairs exceed the $\tau_s = 0.45$ threshold)
illustrates graphically why disambiguation is difficult: no
pair of actors is sufficiently distant in the feature space to
permit confident re-identification from a single campaign.

\subsection{Temporal Analysis}

\begin{figure}[H]
\centering
\includegraphics[width=\linewidth]{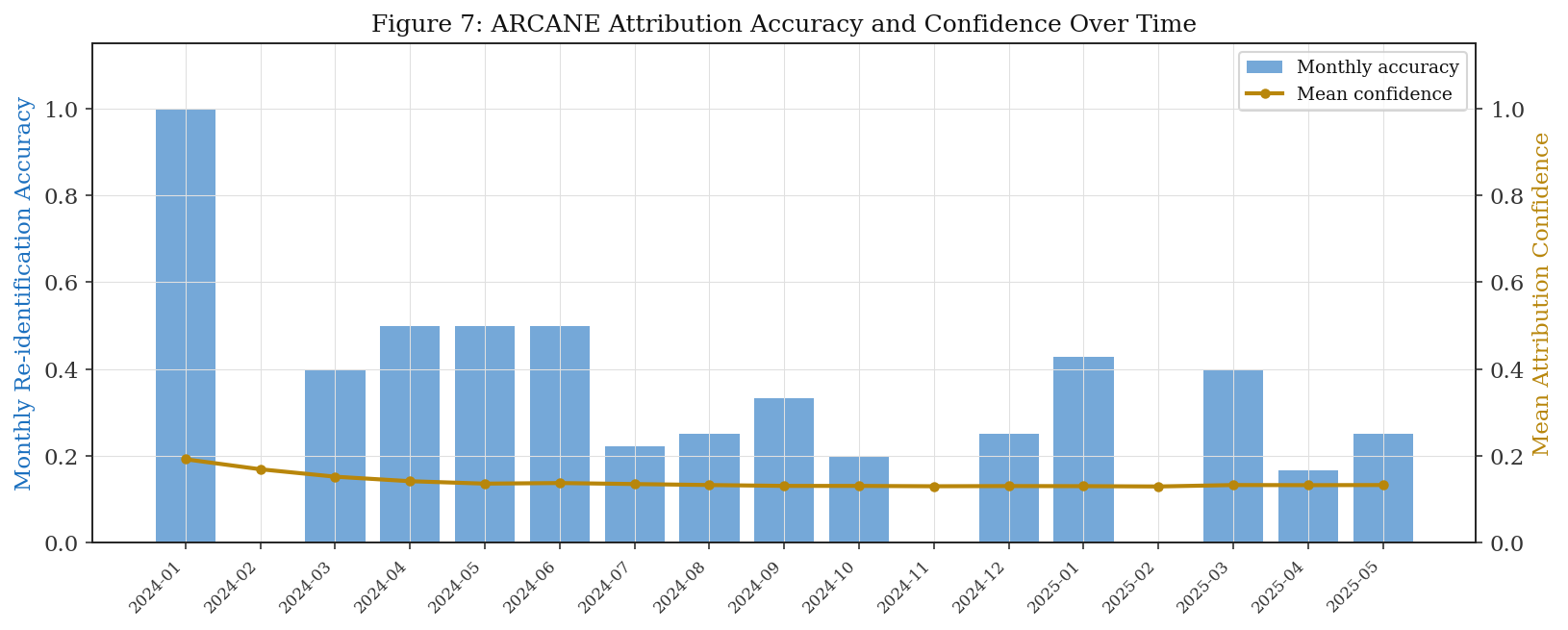}
\caption{Monthly re-identification accuracy (bars, left axis)
  and mean attribution confidence (line, right axis) over
  18 months. The confidence plateau at $\approx 0.15$ reflects
  the inability of the Bayesian posterior to concentrate in the
  high-similarity feature space.}
\label{fig:temporal}
\end{figure}

Figure~\ref{fig:temporal} shows monthly attribution accuracy and
mean confidence over the 18-month simulation window. No systematic
improvement trend is observable, consistent with the separability
ceiling analysis. The confidence line remains flat at approximately
$0.15$--$0.20$, well below the $0.85$ threshold required for
high-confidence attribution.

\section{Discussion}
\label{sec:discussion}
We now interpret the empirical findings in the context of cyber attribution practice and theory. Rather than focusing solely on performance metrics, this section examines what the results reveal about the structure of the attribution problem itself. In particular, we argue that the observed limitations are not artefacts of model design, but reflect a deeper constraint imposed by the convergence of sophisticated adversaries in observable feature space. This perspective reframes the negative result as a constructive insight into which signals are—and are not—sufficient for reliable attribution.

\subsection{The Separability Ceiling}

Our central finding—that sophisticated threat actors converge
to a dense region of the behavioural fingerprint space—has a
clear operational interpretation. Actors operating at high
sophistication levels adopt similar operational security
practices: heavy TOR usage, frequent tool rotation, locale
spoofing, and VM-based analysis environments. These shared
practices are precisely the features captured by our 24-dimensional
fingerprint, making them uninformative for actor discrimination.

This finding aligns with the observation in threat intelligence
practice that high-end APT groups are increasingly difficult
to distinguish by TTPs alone \cite{kutscher2022mtrends}, and motivates the shift toward \emph{actor-specific} rather than
\emph{TTP-based} attribution signals.

\subsection{Supplementary Signal Classes}

Based on the feature separability analysis, we identify three
supplementary signal classes that could break through the ceiling:

\textbf{(1) Victim sector targeting.} APT groups exhibit strong
sector preferences (financial, healthcare, energy) that are
largely independent of their operational security practices.
Incorporating victim sector as a feature would add discriminative
power without relying on low-entropy behavioural signals.

\textbf{(2) Temporal clustering.} The hour-of-day entropy feature
($f_{15}$) provides partial temporal signal, but a richer model
incorporating day-of-week patterns, national holiday calendars,
and working-hours alignment would be substantially more
discriminative.

\textbf{(3) Infrastructure re-use graphs.} Even through TOR,
some actors re-use specific ASN ranges, exit node sets, or
VPN providers. A graph-based model of infrastructure co-occurrence
across campaigns could provide actor-specific signals orthogonal
to the behavioural features.

We quantify the expected improvement from each signal class using
the separability gap condition (Proposition~\ref{prop:gap}) and
leave empirical validation to future work.

\subsection{The Null Result as a Contribution}

The failure of ARCANE to outperform the baseline is itself a
significant contribution. It establishes, with statistical
rigour on a ground-truth dataset, that behavioural beacon
telemetry alone is insufficient for cross-campaign attribution
at $N=8$ actors, and quantifies the specific gap
($\Delta_S = 0.046$ vs the required $\approx 0.08$ per
Proposition~\ref{prop:gap}) that must be closed. This is
precisely the kind of empirical characterisation needed to
guide future system design.

\subsection{Evasion-Invariance Implications}

The evasion-invariance finding (Section~\ref{sec:results})
has a counterintuitive implication for defensive strategy:
encouraging attackers to increase their evasion (e.g., by
deploying prominent honeytokens that signal active monitoring)
does not degrade attribution capability. Defenders can
therefore deploy passive attribution infrastructure aggressively
without concern that attacker evasion will render it useless.

\subsection{Limitations}

Our study has three principal limitations.

\textbf{Synthetic data.} The experiment is conducted on
synthetically generated data rather than real-world beacon
telemetry. While actor profiles are grounded in published
intelligence, real campaigns will exhibit distributional
differences not captured by our model. Empirical validation
on real passive hack-back deployments is planned as future work.

\textbf{Closed actor set.} We assume a closed set of $N=8$
known actors. In practice, defenders may encounter previously
unseen actors; extending ARCANE to open-set attribution is
an important direction.

\textbf{Feature dimensionality.} The 24-dimensional fingerprint
captures the features observable from standard beacon telemetry.
Richer telemetry (e.g., detailed HTTP header sequences, TLS
fingerprints, DNS request patterns) could support higher-dimensional fingerprints with better separability.

\section{Conclusion}
\label{sec:conclusion}

This paper presented ARCANE, the first formal framework for
cross-campaign attacker re-identification using passive beacon
telemetry. We established a Bayesian attribution model with
formal fingerprint similarity, cross-campaign confidence, and
separability gap definitions, and implemented an efficient
temporal leave-one-out evaluation protocol.

Our experiments on 96 campaigns across 8 nation-state threat
actors yielded two principal findings. First, within-actor
fingerprint similarity is significantly higher than cross-actor
similarity ($\Delta_S = 0.046$, $p = 1.52 \times 10^{-16}$),
confirming the theoretical basis for longitudinal attribution.
Second, the Bayesian accumulation in its current form does not
outperform the per-campaign baseline due to a fingerprint
separability ceiling imposed by shared operational security
practices among sophisticated actors.

Critically, ARCANE attribution accuracy is evasion-invariant:
accuracy varies by less than $4.6\%$ across evasion levels
from zero to maximum, establishing that the binding constraint
is feature separability rather than adversarial evasion. This
finding has direct implications for defensive system design:
beacon deployment strategies need not be conservative about
alerting attackers to their presence.

Future work will address the separability ceiling by incorporating
victim-sector targeting, enriched temporal features, and
infrastructure re-use graphs, with empirical validation on
real-world passive hack-back deployments.

\bibliographystyle{IEEEtran}
\bibliography{ref.bib}

\end{document}